\documentclass[amssymb,amsmath,aps,secnumarabic,floatfix,showpacs]{revtex4}
\usepackage{psfig}
\usepackage{epsfig}
\usepackage{pstricks}
\usepackage{dcolumn}
\newpsobject{showgrid}{psgrid}{subgriddiv=1}
\expandafter\ifx\csname package@font\endcsname\relax\else
 \expandafter\expandafter
 \expandafter\usepackage
 \expandafter\expandafter
 \expandafter{\csname package@font\endcsname}
\fi
\begin{document}
\title{Asynchronously parallelised percolation on distributed machines}
\author{Nicholas R. Moloney}
\email{n.moloney@imperial.ac.uk}
\affiliation{
Blackett Laboratory,
Imperial College London,
Prince Consort Road,
London SW7 2BW,
United Kingdom}
\author{Gunnar Pruessner}
\email{gunnar.pruessner@physics.org}
\affiliation{
Department of Mathematics,
Imperial College London,
180 Queen's Gate,
London SW7 2BZ,
United Kingdom}
\date{\today}

\begin{abstract}
We propose a powerful method based on the Hoshen-Kopelman algorithm
for simulating percolation asynchronously on distributed machines.  
Our method demands very little of hardware and yet we are able to make high
precision measurements on very large lattices.  We implement our method
to calculate various cluster size distributions on large lattices of
different aspect ratios spanning three orders of magnitude for two-dimensional site 
and bond percolation. We find that the nonuniversal
constants in the scaling function for the cluster size distribution apparently satisfy
a scaling relation, and that the moment ratios for the largest cluster size distribution 
reveal a characteristic aspect ratio at $r \approx 9$.
\end{abstract}
\maketitle

Although an old problem \cite{Flory:1941}, percolation continues to
attract a steady stream of papers
\cite{Duplantier:1999,AizenmanDuplantierAharony:1999,NewmanZiff:2000}.
High-quality numerical data are required to corroborate the many
analytical results, particularly from conformal field theory
\cite{Cardy:1992,LanglandsPichetPouliotStAubin:1992,Pinson:1994,Cardy:1998}.
In this paper, we describe a method of simulating percolation that runs
asynchronously in parallel on almost any hardware.  In principle, the
method relaxes all the standard constraints in numerical simulations
of percolation, such as CPU power, memory, and network capacity.  It
is especially suited for calculating cluster size distributions,
finite size corrections, crossing probabilities, and, by
applying the corresponding boundary conditions, distributions of
wrapping clusters on different topologies, e.g., cylinder, torus, or
the M\"{o}bius strip.

The Hoshen-Kopelman algorithm (HKA) \cite{HoshenKopelman:1976} is still
the standard technique for identifying clusters in percolation, 
where a cluster is a set of sites connected via nearest neighbour 
interactions (site percolation) or active bonds (bond percolation). 
Strictly speaking, it is a type of data representation particularly
suited for tracking clusters.  Recently, Newman and Ziff
\cite{NewmanZiff:2000} have shown how to exploit this data
representation to monitor the change in various observables as the
occupation probability $p$ is increased.  The data representation
efficiently encodes the connectivity of clusters in a large percolation
system. In this paper, we show how to exploit this representation for
different system sizes (up to $\approx 5 \times 10^{14}$ sites) and aspect 
ratios. The algorithm runs asynchronously in parallel over an almost 
arbitrarily slow network of computers.  The network is hierarchically 
organised, and nodes on lower levels (slave nodes) can be slow 
and heterogeneous.  In fact, the system scales like an ideal parallel 
computer: the overall computing time decreases linearly with the number 
of nodes, especially for large slave lattices (patches) where the overhead due to 
networking and related processing and the CPU-time at higher levels 
(master nodes) becomes negligible (see Tab.~\ref{tab:overhead}). Other
memory-efficient methods exist for constructing large clusters, for example
Paul, Ziff, and Stanley \cite{PaulZiffStanley:2001}. In that paper a variant of the 
Leath algorithm is used \cite{Leath:1976}, together with a data structure to record
information about visited sites. As a result, memory is made available as and 
when it is required. In our method we can easily count the number of spanning
clusters per realisation, apply different boundary conditions, rearrange patches
for different aspect ratios, and gather statistics at every stage of lattice 
construction. 

We describe our method in detail for two-dimensional site percolation on
a square lattice and present the overall cluster size distribution for
different aspect ratios, as well as the universal moment ratios for the
distribution of the order parameter in site and bond percolation. 
We find two surprising results. First, the nonuniversal amplitudes 
in the scaling function for the cluster number distribution numerically 
satisfy a scaling relation. Second, the moment ratios for the largest 
cluster size distribution all peak at $r \approx 9$, 
defining a characteristic aspect ratio.

The basic idea of the method is that many slave nodes independently
simulate lattices of equal linear size $L$ in parallel using the
HKA.  These nodes send a special representation of their lattice border
to a master node, which combines $m$ of these patches to form a superlattice.
The advantage of such a decomposition is that the master node can build
up a very large superlattice while maintaining the large scale
histograms.  The master node can apply different boundary conditions and
even reuse the same patches several times by rotating, mirroring, and
permuting them.  

The key to our algorithm is the representation of the lattice borders
by the slave nodes. This is essentially a form of path compression (or
Nakanishi label recycling \cite{BinderStauffer:1987}), where all border
sites are considered active (i.e., possibly changing connectivity) and 
bulk sites are considered inactive. In this representation information about 
the connectivity and size of any cluster connected to the border is
summarised entirely within border sites. The spatial information of
clusters is neither required nor stored. Thus clusters not connected to
the border are ignored, although their contribution to the cluster size
histogram is recorded locally.  

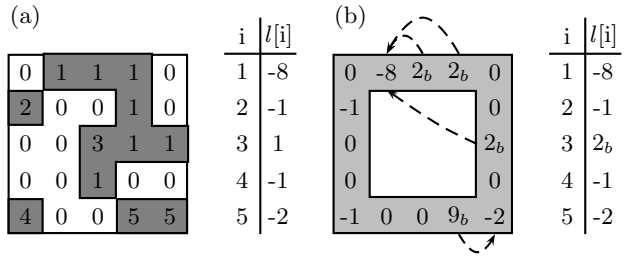
\begin{figure}[ht]
\begin{center}
\psset{xunit=0.48cm,yunit=0.48cm}
\begin{pspicture}(3,0)(16,6)

\psframe(1,0)(6,5)
\pspolygon[fillstyle=solid,fillcolor=gray](2,4)(4,4)(4,3)(3,3)(3,1)(4,1)(4,2)(6,2)(6,3)(5,3)(5,5)(2,5)
\psframe[fillstyle=solid,fillcolor=gray](1,3)(2,4)
\psframe[fillstyle=solid,fillcolor=gray](1,0)(2,1)
\psframe[fillstyle=solid,fillcolor=gray](4,0)(6,1)

\rput(1.5,4.5){0} \rput(2.5,4.5){1} \rput(3.5,4.5){1} \rput(4.5,4.5){1} \rput(5.5,4.5){0} 
\rput(1.5,3.5){2} \rput(2.5,3.5){0} \rput(3.5,3.5){0} \rput(4.5,3.5){1} \rput(5.5,3.5){0} 
\rput(1.5,2.5){0} \rput(2.5,2.5){0} \rput(3.5,2.5){3} \rput(4.5,2.5){1} \rput(5.5,2.5){1} 
\rput(1.5,1.5){0} \rput(2.5,1.5){0} \rput(3.5,1.5){1} \rput(4.5,1.5){0} \rput(5.5,1.5){0} 
\rput(1.5,0.5){4} \rput(2.5,0.5){0} \rput(3.5,0.5){0} \rput(4.5,0.5){5} \rput(5.5,0.5){5} 

\psline(8,6)(8,0) \rput(7.5,5.5){i}
\psline(7,5)(9,5) \rput(8.5,5.5){\textit{l}[i]}

\rput(7.5,4.5){1} \rput(8.5,4.5){-8}	
\rput(7.5,3.5){2} \rput(8.5,3.5){-1}	
\rput(7.5,2.5){3} \rput(8.5,2.5){1}	
\rput(7.5,1.5){4} \rput(8.5,1.5){-1}	
\rput(7.5,0.5){5} \rput(8.5,0.5){-2}	

\psframe[fillstyle=solid,fillcolor=lightgray](10,0)(15,5)
\psframe[fillstyle=solid,fillcolor=white](11,1)(14,4)
\rput(10.5,4.5){0} \rput(11.5,4.5){-8} \rput(12.5,4.5){$2_b$} \rput(13.5,4.5){$2_b$} \rput(14.5,4.5){0} \rput(14.5,3.5){0} \rput(14.5,2.5){$2_b$} \rput(14.5,1.5){0}
\rput(14.5,0.5){-2} \rput(13.5,0.5){$9_b$} \rput(12.5,0.5){0} \rput(11.5,0.5){0} \rput(10.5,0.5){-1} \rput(10.5,1.5){0} \rput(10.5,2.5){0} \rput(10.5,3.5){-1}
\pscurve[linestyle=dashed]{->}(14,2.5)(13,3.0)(11.5,4)
\pscurve[linestyle=dashed]{->}(12.5,5)(12,5.5)(11.5,5)
\pscurve[linestyle=dashed]{->}(13.5,5)(12.5,6)(11.5,5)
\pscurve[linestyle=dashed]{->}(13.5,0)(14,-0.5)(14.5,0)

\psline(17,6)(17,0) \rput(16.5,5.5){i}
\psline(16,5)(18,5) \rput(17.5,5.5){\textit{l}[i]}

\rput(16.5,4.5){1} \rput(17.5,4.5){-8}	
\rput(16.5,3.5){2} \rput(17.5,3.5){-1}	
\rput(16.5,2.5){3} \rput(17.5,2.5){$2_b$}	
\rput(16.5,1.5){4} \rput(17.5,1.5){-1}	
\rput(16.5,0.5){5} \rput(17.5,0.5){-2}

\rput(1.5,6){(a)} \rput(10.5,6){(b)}

\end{pspicture}
\caption{ \label{fig:border_scan}(a) The lattice and the list of labels, $l[i]$, 
as prepared by the HKA. (b) The border configuration suitable 
for the master node, after a clockwise border scan starting in the upper left-hand corner, 
with the list of labels now being irrelevant. 
For the reader's convenience, labels pointing to sites $i$
in the new border carry a suffix $i_b$. } 
\end{center}
\end{figure}
The HKA produces a list of labels, to which all active sites refer in
order to identify their cluster, see Fig.~\ref{fig:border_scan}(a).  
After the realisation of a lattice, a \emph{new} border representation 
is prepared by visiting each border site in succession, 
indexed from $1$ to $4L-4$, see Fig.~\ref{fig:border_scan}(b). 
The first site of a previously unscanned cluster contains the size 
of the cluster as a negative value in the range $[-1, -L^2]$. 
This site is called the root. In the list of labels of the original 
representation, the label of this cluster is changed to indicate 
the new location of the root site in the border. 
All other sites in the border which belong to the same cluster
refer to this site. The slave nodes send the border configuration
in this representation to the master node.  If required,
clusters in the bulk have their sizes recorded in a local histogram,
i.e. at the slave node that produced the lattice. This histogram is
stored locally for the duration of the simulation. The master node is
the only component that requires enough memory to hold the large
histogram(s) usually generated in large scale simulations, while the
slaves only need to store a very small amount of local data.

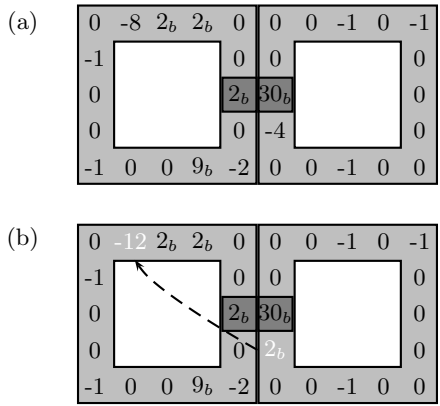
\begin{figure}[ht]
\begin{center}
\psset{xunit=0.48cm,yunit=0.48cm}
\begin{pspicture}(0,0)(15,6)

\psframe[fillstyle=solid,fillcolor=lightgray](2,0)(7,5)
\psframe[fillstyle=solid,fillcolor=white](3,1)(6,4)
\psframe[fillstyle=solid,fillcolor=lightgray](7,0)(12,5)
\psframe[fillstyle=solid,fillcolor=white](8,1)(11,4)
\psframe[fillstyle=solid,fillcolor=gray](7,2)(8,3)
\psframe[fillstyle=solid,fillcolor=gray](6,2)(7,3)

\rput(2.5,4.5){0} \rput(3.5,4.5){-8} \rput(4.5,4.5){$2_b$} \rput(5.5,4.5){$2_b$} \rput(6.5,4.5){0} 
\rput(6.5,3.5){0} \rput(6.5,2.5){$2_b$} \rput(6.5,1.5){0} \rput(6.5,0.5){-2} 
\rput(5.5,0.5){$9_b$} \rput(4.5,0.5){0} \rput(3.5,0.5){0} \rput(2.5,0.5){-1} 
\rput(2.5,1.5){0} \rput(2.5,2.5){0} \rput(2.5,3.5){-1}

\rput(7.5,4.5){0} \rput(8.5,4.5){0} \rput(9.5,4.5){-1} \rput(10.5,4.5){0} \rput(11.5,4.5){-1}
\rput(11.5,3.5){0} \rput(11.5,2.5){0} \rput(11.5,1.5){0} \rput(11.5,0.5){0}
\rput(10.5,0.5){0} \rput(9.5,0.5){-1} \rput(8.5,0.5){0} \rput(7.5,0.5){0}
\rput(7.5,1.5){-4} \rput(7.5,2.5){$30_b$} \rput(7.5,3.5){0}

\rput(0.5,4.5){(a)} \rput(0.5,-1.5){(b)}

\end{pspicture}

\begin{pspicture}(0,0)(15,6)

\psframe[fillstyle=solid,fillcolor=lightgray](2,0)(7,5)
\psframe[fillstyle=solid,fillcolor=white](3,1)(6,4)
\psframe[fillstyle=solid,fillcolor=lightgray](7,0)(12,5)
\psframe[fillstyle=solid,fillcolor=white](8,1)(11,4)
\psframe[fillstyle=solid,fillcolor=gray](7,2)(8,3)
\psframe[fillstyle=solid,fillcolor=gray](6,2)(7,3)


\rput(2.5,4.5){0} {\white \rput(3.5,4.5){-12} } \rput(4.5,4.5){$2_b$} \rput(5.5,4.5){$2_b$} \rput(6.5,4.5){0} 
\rput(6.5,3.5){0} \rput(6.5,2.5){$2_b$} \rput(6.5,1.5){0} \rput(6.5,0.5){-2}
\rput(5.5,0.5){$9_b$} \rput(4.5,0.5){0} \rput(3.5,0.5){0} \rput(2.5,0.5){-1}
\rput(2.5,1.5){0} \rput(2.5,2.5){0} \rput(2.5,3.5){-1}

\rput(7.5,4.5){0} \rput(8.5,4.5){0} \rput(9.5,4.5){-1} \rput(10.5,4.5){0} \rput(11.5,4.5){-1}
\rput(11.5,3.5){0} \rput(11.5,2.5){0} \rput(11.5,1.5){0} \rput(11.5,0.5){0}
\rput(10.5,0.5){0} \rput(9.5,0.5){-1} \rput(8.5,0.5){0} \rput(7.5,0.5){0}
\rput(7.5,2.5){$30_b$} \rput(7.5,3.5){0}
\white{\rput(7.5,1.5){$2_b$}} 

\pscurve[linestyle=dashed]{->}(7,1.5)(4,3.5)(3.6,4)

\end{pspicture}

\end{center}
\caption{ \label{fig:join_border} 
(a) The configuration of the borders before two clusters merge at
the marked labels. The labels in the right patch are shifted by
$4L-4$ to make them unique. (b) The configuration of the borders
after the merging procedure. Labels which have changed are shown in
white.  }
\end{figure}

When two patches are combined by the master (gluing) it is possible
that two clusters merge at the border. This is realised by
setting one of the root labels (preferably from the smaller cluster) to point to the other, 
as shown in Fig.~\ref{fig:join_border}. The master's histogram is updated by
removing both cluster sizes ($4$ and $8$ in the example) and replacing
them by their sum.  
By adding the site-normalised histogram of the slaves (i.e., the number density of
$s$-clusters), $n_\slave(s)$, to the site-normalised histogram on the master node, 
$n_\master(s)$, the total histogram, $n(s)$ is obtained,
\begin{equation}
 	n(s) = n_\master(s) + n_\slave(s) \ . \label{eq:superimposed_histos}
\end{equation}
This result does not involve any approximation and is independent of
the number of realisations. Because the
superposition Eq.~\eqref{eq:superimposed_histos} can be postponed
until postprocessing, the slaves can store these data locally.
Moreover, because all relevant information is encoded in the patches,
when and whence they arrive at the master node is arbitrary.  Hence
the algorithm is asynchronous, in contrast to standard techniques of
parallelisation, for example Ref.~\cite{Tiggemann:2001}.

The master node can itself be considered a slave node and prepare a
border configuration for another master node, so that one obtains a
tree-like structure of master and slave nodes, where statistics can be
obtained on every level. We have used this scheme to produce a single lattice of size 
$(22.2\times 10^6)^2$ sites, and have calculated its cluster size distribution.  
For large $L$ the CPU-time required for networking becomes negligible, 
as shown in Table~\ref{tab:overhead}. The complexity of the master gluing algorithm 
is $\OC(m L \log L)$, while the slaves need $\OC(L^2 \log L)$ time to produce a patch, 
which is represented in $\OC(L)$ memory. Therefore, the optimal number of slaves per 
master in which the master fully utilises its resources, 
while not blocking any slaves, scales like $L$. At the same time, the relative 
networking overhead per slave scales like $1/L$. Table~\ref{tab:overhead} 
shows the corresponding measurements.

\begin{table}[ht]
\begin{ruledtabular}
\newcolumntype{d}[1]{D{.}{.}{#1}}
\begin{tabular}{ccd{3}}
L      & Slave nodes per master & \multicolumn{1}{c}{Approximate overhead} \\ \hline
100    & 2                      & 4.8\% \\
200    & 4                      & 2.9\% \\
500    & 10                     & 1.4\% \\
1000   & 22                     & 1.7\%
\end{tabular}
\end{ruledtabular}
\caption{The optimal number of slaves and relative networking overhead of
the slave nodes. The master node used was roughly twice as fast as the
slave nodes and applied six different boundary conditions on $14$
different aspect ratios from each set of $900$ patches of size
$L^2$ produced by the slaves. 
\label{tab:overhead}
}
\end{table}

Debarring correlations introduced by the random number generator, all
patches arriving at the master are statistically independent.  
However, it is possible to recycle incoming patches by arranging them in
different configurations (e.g., boundary conditions or aspect ratios). 
The results for these different configurations are {\it not}
statistically independent. An upper bound can be calculated for the
error introduced by this procedure. Rather than recycling all patches
$q$ times (e.g., for $q=14$ different aspect ratios), one could 
distribute them evenly among $q$ bins, now all statistically independent.  
The error in the estimator for the mean of an observable in the $q$ bins 
would be a factor $\sqrt{q}$ larger than that for the complete sample.  
Therefore, when considering results for $q$ bins while using the same complete 
sample in each bin, the upper bound for the error is $\sqrt{q}$ times the error for 
the complete set. When patches are recycled it is possible to reduce the correlations 
by randomly rotating, mirroring and permuting them.
We have done this in all simulations.
\\
\\
As an application of the algorithm, various cluster size distributions
for site and bond percolation for $q=14$ different aspect ratios, $r=$ width/height, 
between $1$ and $900$ were calculated. 
The slaves produced square patches of three different sizes, 
$L=10,100,1000$, of which $m = 900$ were glued at the master node 
to form $q$ superlattices with $N=300^2,3000^2,30000^2$ sites. 
The simulations were performed at critical density $p_c=0.59274621$ for site percolation 
\cite{NewmanZiff:2000}, and $p_c=1/2$ for bond percolation \cite{Kesten:1980}.
All numerical results are based on at least $10^6$ independent realisations 
(i.e., roughly $10^9$ realisations at the slave nodes). Free boundary conditions 
have been applied everywhere. The random number generator used was the so-called 
Mersenne-Twister \cite{MatsumotoNishimura:1998b},
which is highly suitable for parallel simulations.

\begin{figure}[ht]
\scalebox{0.5}{\includegraphics{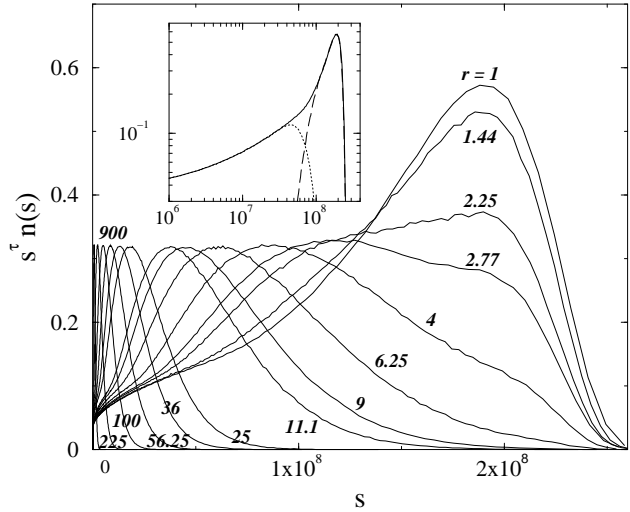}}
\caption{The rescaled and binned distribution $s^\tau n_s(s)$ for systems containing
$N=30000^2$ sites. 
The inset shows $s^{\tau}n_s(s)$ (solid line), $s^{\tau}n_{\text{max};s}(s)/N$, (long-dashed line), 
and their difference (dotted line), for $r = 1$. 
Evidently, the bump in the distribution is derived mainly from the largest clusters.  
\label{fig:histo_all}
}
\end{figure}

The site-normalised cluster size distribution $n_{s,b}(s;r)$ is the number density 
of $s$-clusters for aspect ratio $r$. Henceforth, subscripts $s$ and $b$ refer 
to site and bond percolation, respectively. For large cluster sizes near $p_c$, 
$n_{s,b}(s;r)$ is expected to behave like
\begin{equation}\label{eq:simple_scaling}
	n_{s,b}(s;r) = a_{s,b}(r) s^{-\tau} \mathcal{G}(s/s^0_{s,b}; r)
\end{equation}
where, in a finite system of effective size $\tilde{L}$, $s^0_{s,b} = b_{s,b}(r) \tilde{L}^D$,
and $\mathcal{G}$ is the scaling function.
The effective size can be taken as anything that scales linearly in $\sqrt{N}$.
The universal critical exponents are $\tau$ and $D$, while the amplitudes 
$a_{s,b}(r)$ and $b_{s,b}(r)$ are nonuniversal, and set by two arbitrary conditions on $\mathcal{G}$. 
Figure~\ref{fig:histo_all} shows 
$s^\tau n_s(s;r)$ for different values of $r$, using $\tau=187/91$ \cite{StaufferAharonyENG:1994}. 

\begin{figure}[th]
\scalebox{0.47}{\includegraphics{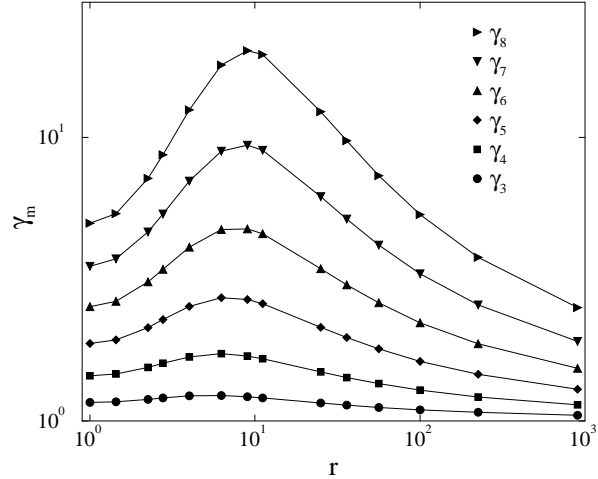}}
\caption{Universal moment ratios $\gamma_{m; s,b}(r)$ for different aspect ratios and system sizes. 
\label{fig:larg_moments}}
\end{figure}

Two interesting features emerge. Independently of $L$, the shape of the
distribution changes abruptly at around $r=2.25$ and the maximum of the
rescaled distribution is seemingly constant for larger $r$. Therefore, there 
is no possible choice of $\tilde{L}$ that can collapse the scaling function for 
different aspect ratios, and $\mathcal{G}$ explicitly depends on $r$. The inset
in Fig.~\ref{fig:histo_all} shows $n_s(s;r)-n_{\li s}(s;r)/N$ at $r=1$, where
$n_{\li s,b}(s;r)$ denotes the distribution of the size of the
largest cluster.  It seems that the sudden change in the shape of
$n_{s,b}(s;r)$ is caused by a change in $n_{\text{\tiny{max;s,b}}}(s;r)$, 
but what happens at this particular value of $r$ remains an open question.

If we define the moment ratios as
\begin{equation}
 V_{m ; s,b}(r) \equiv \frac{\langle s^m \rangle_{s,b}(r) N}{\langle s^2 \rangle^{m/2}_{s,b}(r) N^{m/2}}
\label{eq:def_uar}
\end{equation}
with $\ave{s^k}_{s,b}(r) \equiv \int s^k n_{s,b}(s;r) \, ds$, 
then site and bond percolation should differ by powers of the factor
\begin{equation}
 	\frac{a_s(r)/a_b(r)}{(b_s(r)/b_b(r))^{\tau - 1}} 
\end{equation}
which is obtained by calculating the moments with the help of Eq.~\eqref{eq:simple_scaling}. 
However, we find numerically that this factor is unity, 
i.e., that the ratio $a(r)/b(r)^{\tau-1}$ is the same 
for site and bond percolation. This ratio is not a universal function, 
because its value depends on the conditions imposed 
on $\mathcal{G}$ for determining $a(r)$ and $b(r)$. However, numerics 
suggests strongly that, once these conditions are given, this ratio is 
independent of the lattice type, i.e., Eq.~\eqref{eq:def_uar} represents a \emph{universal} moment ratio. 
Therefore, it is possible to write 
\begin{equation} 
   a_{s,b}(r) = b_{s,b}^{\tau-1}(r) q(r), 
  \label{eq:defq} 
\end{equation} 
where $q(r)$ depends only on the choice of the two conditions imposed 
on $\mathcal{G}$, but not on the lattice type. As mentioned above, $\mathcal{G}$ is necessarily an 
explicit function of $r$, so that it can absorb $q(r)$ defined in 
Eq.~\eqref{eq:defq}, thereby fixing one of the two conditions on $\mathcal{G}$. 
The remaining condition determines (together with the 
choice of $\tilde{L}$) the remaining free parameter. 
Consequently, we conclude that Eq.~\eqref{eq:simple_scaling} 
can be replaced by 
\[ 
   n_{s,b}(s;r) = b_{s,b}^{\tau-1}(r) s^{-\tau} 
            \widetilde{\mathcal{G}}(s/(N^{D/2} b_{s,b}); r). 
\] 
Of course, $b_{s,b}(r)^{\tau-1}$ cannot be absorbed into $\mathcal{G}$ in the same way as $q(r)$ because it depends on the lattice type.
Thus all the characteristics of the lattice enter solely through $b$.
For completeness we note that numerically the ratios $a_s(r)/a_b(r)$ and $b_s(r)/b_b(r)$ 
are independent of $r$, no matter what conditions are imposed on $\mathcal{G}$.

The order parameter of percolation is the fraction of sites belonging to
the spanning (or largest) cluster. Thus, one expects the moment ratios
\begin{equation}
	\gamma_{m} \equiv \frac{\ave{s^m}_{\li s,b}(r)}{\ave{s^2}^{m/2}_{\li s,b}(r)}
\end{equation}
with $\ave{s^k}_{\li s,b}(r) \equiv \int s^k n_{\li s,b}(s;r) \, ds$
to be universal. Figure~\ref{fig:larg_moments} shows the behavior of this 
ratio for different aspect ratios $r$. A pronounced bump appears at around $r=9$. 
The origin of the bump remains unclear, and can be used to define a characteristic aspect ratio.

In conclusion, the method proposed in this paper permits the use of 
resources usually considered too slow, small, or badly connected. 
At the same time, it takes advantage of parallelisation by providing a very 
flexible framework for simulating different boundary conditions and aspect ratios.
By way of illustration, we have increased Tiggemann's former world record 
\cite{Tiggemann:2001} for the largest simulated system by a factor of $30$. 
The new record was set by an undergraduate computer cluster (as opposed to a Cray T3E)
when idle. The data presented have remarkable numerical accuracy and are 
from systems of unprecedented size. They give rise to a number
of urgent questions, namely how to reconsider the non-universal 
amplitudes in Eq.~\eqref{eq:simple_scaling}, and how to account
for a characteristic aspect ratio as provided by the moment ratios of the
largest cluster size distribution. 

\begin{acknowledgments}
The authors wish to thank Andy Thomas for his fantastic technical
support.  Without his help and dedication, this project would not have
been possible.  The authors are grateful for the generous donation of
``I-D Media AG, Application Servers \& Distributed
Applications Architectures, Berlin''.
We especially thank Matthias Kaulke and Oliver Kilian.
The authors also thank Dan Moore, Brendan Maguire, and Phil Mayers for their continuous
support, as well as Kim Christensen for his helpful comments. 
N.R.M. is very grateful to the Beit Fellowship, and to the Zamkow family. 
G.P. gratefully acknowledges the support of the EPSRC.
\end{acknowledgments}
\bibliography{articles,books}
\end{document}